\documentclass[10pt,a4paper]{article}
\pdfoutput=1
\usepackage{jcappub}

\newcommand{\newc}{\newcommand}
\newc{\beq}    {\begin{equation}}
\newc{\eeq}    {\end{equation}}

\newc{\beqa}    {\begin{eqnarray}}
\newc{\eeqa}    {\end{eqnarray}}
\newc{\bs}    {\section}
\newc{\no}    {\\ \nonumber}

\newcommand{\bea}{\begin{eqnarray}}
\newcommand{\eea}{\end{eqnarray}}

\newc{\st}    {\stackrel}

\title{Cosmology of self-interacting ultralight dark matter: simultaneous shifts in the Hubble constant and structure growth}

\author[a]{Seokcheon Lee}
\author[b]{and Jae-Weon Lee}
\affiliation[a]{Department of Physics, Institute of Basic Science, Sungkyunkwan University, Suwon 16419, Korea}
\affiliation[b]{Department of Electrical and Electronic Engineering, Jungwon University, 85 Munmu-ro, Goesan-eup, Goesan-gun, Chungcheongbuk-do 28024, Korea}
\emailAdd{skylee@skku.edu}
\emailAdd{scikid@jwu.ac.kr}

\abstract{We study the cosmology of self-interacting ultralight dark matter. The scalar field evolves from an early Hubble-frozen phase to a quartic-dominated radiation-like phase and finally to a quadratic matter-like phase. The intermediate phase increases the pre-recombination expansion rate and reduces the sound horizon, while the late phase allows the same field to contribute to dark matter. We implement the background and linear perturbations of a mixed cold-dark-matter and ultralight-dark-matter cosmology in CLASS. For $m=10^{-22}\,{\rm eV}$ and $\tilde m\simeq1\,{\rm eV}$, increasing the ultralight-dark-matter fraction raises the fixed-$\theta_*$ value of $H_0$, lowers the late-time clustering amplitude, strengthens the scale-dependent matter-power suppression, and produces a smooth CMB-temperature residual reaching about $9\%$ for the largest fraction considered. The model therefore generates simultaneous background and growth shifts in the directions relevant to the Hubble and structure-growth tensions. Establishing whether these shifts improve a joint fit requires a likelihood-level observational analysis.}


\begin{document}
\maketitle
\flushbottom

\section{Introduction}
\label{sec:introduction} 

The standard $\Lambda$CDM model successfully accounts for a wide range of cosmological observations, but persistent discrepancies remain between early- and late-Universe measurements. The most prominent is the Hubble tension, namely the difference between the locally measured value of the present expansion rate and the value inferred from the cosmic microwave background (CMB) within $\Lambda$CDM~\cite{DiValentino:2021izs,Kamionkowski:2022pkx,Wang:2023bxf,Verde:2023lmm}. A related, though less uniform, discrepancy concerns the amplitude of late-time matter clustering, commonly parametrized by $S_8\equiv\sigma_8\sqrt{\frac{\Omega_m}{0.3}}$, for which several weak-lensing and galaxy-clustering analyses have obtained central values below the Planck $\Lambda$CDM prediction~\cite{Planck:2018vyg,KiDS:2020suj,DES:2021wwk,Sugiyama:2023fzm}.

The positions of the CMB acoustic peaks constrain primarily the angular sound-horizon scale, $\theta_* = \frac{r_s(z_*)}{D_M(z_*)}$, rather than $H_0$ directly.  A reduction of the sound horizon $r_s(z_*)$, caused by an enhanced pre-recombination expansion rate, can therefore allow a larger value of $H_0$ while preserving the accurately measured value of $\theta_*$.  This observation motivates early-time solutions involving additional radiation-like energy density, early dark energy, or oscillating scalar fields~\cite{Flambaum:2019cih,Jedamzik:2020zmd,Gonzalez:2020fdy,Ye:2021iwa}. However, a viable model must also remain consistent with the growth of cosmic structure.  In particular, an early-time modification that raises the fixed-$\theta_*$ value of $H_0$ should not simultaneously worsen the discrepancy in the late-time clustering amplitude.

A closely related phenomenology was obtained in Ref.~\cite{Ye:2021iwa}, where an AdS early-dark-energy (EDE) field reduces the sound horizon and a separate ultralight axion component suppresses late-time clustering. The present model differs in that both effects arise from successive dynamical regimes of a single self-interacting scalar field: its quartic-dominated oscillatory phase is radiation-like, whereas its subsequent quadratic phase is matter-like. No additional EDE sector is required. The mechanism also differs from Rock~'n'~Roll early dark energy~\cite{Agrawal:2019lmo}: here the radiation-like episode is a phase of the dark matter (DM) field itself, which subsequently becomes matter-like, rather than a separate component introduced only for the early-time expansion history.

Ultralight scalar fields provide a natural framework in which both the background expansion and structure formation can be modified. In the quadratic regime, an oscillating ultralight field behaves as pressureless matter on scales larger than its effective Jeans scale, while wave pressure suppresses clustering on sufficiently small scales~\cite{Baldeschi:1983mq,Sin:1992bg,Lee:1995af,Matos:1998vk,Hu:2000ke,Boehmer:2007um,Marsh:2010wq,Hlozek:2014lca}. These properties have motivated ultralight dark matter (ULDM) as an alternative to conventional cold dark matter (CDM), although the simplest noninteracting models are strongly constrained by observations of small-scale structure~\cite{Irsic:2017yje,Armengaud:2017nkf,Rogers:2020ltq}. ULDM has also been investigated in connection with planar satellite distributions and the final-parsec evolution of supermassive-black-hole binaries~\cite{Park:2022lel,Koo:2023gfm,Bromley:2023yfi}; broader reviews of its galactic and cosmological phenomenology are given in Refs.~\cite{Hui:2016ltb,Ferreira:2020fam,Matos:2023usa}.

A repulsive quartic self-interaction qualitatively changes the cosmological evolution of the field.  For
\beq
V(\phi) = \frac{1}{2}m^2\phi^2 + \frac{\lambda}{4}\phi^4, \qquad \lambda>0, \label{eq:intro-potential}
\eeq
the field evolves through three successive regimes. At early times, it is frozen by Hubble friction and has $w_\phi\simeq-1$. Once oscillations begin while the quartic term dominates, the oscillation-averaged EOS is $\langle w_\phi\rangle=1/3$, so that the field behaves as an additional radiation-like component. As the oscillation amplitude decreases, the quadratic term eventually dominates, and the field approaches $\langle w_\phi\rangle=0$, thereby becoming matter-like.

This mechanism was proposed in Ref.~\cite{Lee:2025roi} as a possible solution to the Hubble tension. A recent phenomenological study examined the correlated response of the sound horizon and late-time structure growth to a transient radiation-like phase in self-interacting ULDM~\cite{Lopez-Sanchez:2025uwk}. By parametrizing the self-interaction contribution as a localized modification of the background expansion, it derived analytic response kernels for the sound horizon and the linear growth amplitude, showing that an episode near matter-radiation equality can produce correlated shifts toward a larger acoustic-scale-compatible $H_0$ value and reduced late-time clustering. That analysis was designed to isolate this phenomenological response and did not evolve the scale-dependent scalar-field perturbations in a Boltzmann solver. The present work complements and extends this approach by evolving the underlying scalar-field background and linear perturbations in CLASS, including the transition from the exact Klein-Gordon (KG) description to an oscillation-averaged effective fluid.

The intermediate quartic phase increases the expansion rate before recombination and reduces the sound horizon, whereas the later quadratic phase allows the same field to contribute to DM. The quartic-to-quadratic transition is characterized by $\tilde m \equiv m / \lambda^{1/4}$. Values near the eV scale place this transition close to matter--radiation equality. For a dominant self-interacting scalar component, representative cosmological and galactic analyses commonly place this combination at an order-eV scale, approximately $1\,{\rm eV}\lesssim\tilde m\lesssim10\,{\rm eV}$~\cite{Li:2013nal,Hartman:2021upg}. As discussed below, the applicable bound is model- and fraction-dependent in a mixed CDM+ULDM cosmology.

However, the background mechanism alone is not sufficient because the cosmological viability of the model depends critically on its perturbations. The scalar perturbations first obey the exact perturbed KG equations and are subsequently represented by an oscillation-averaged effective fluid after many resolved cycles. Previous studies have shown that this handoff must consistently transfer the averaged stress-energy and metric sources~\cite{Urena-Lopez:2015gur,Cookmeyer:2019rna,Passaglia:2022bcr,Liu:2024yne,Moss:2025ymr}. In the quadratic regime, the effective sound speed vanishes as $k\rightarrow0$, so the scalar must recover the pressureless growing mode on large scales. Preserving the scalar momentum density across the KG-to-fluid handoff is essential for distinguishing physical Jeans-scale suppression from numerical loss of matter power.

In this work, we develop the background and linear-perturbation cosmology of self-interacting ULDM and implement it in CLASS~\cite{Lesgourgues:2011re,Blas:2011rf}. We use the mixed dark sector defined in Sec.~\ref{subsec:mixed-background-hubble}, with a present-day scalar fraction that continuously controls the ULDM contribution. The exact KG system is evolved through the physical onset of rapid oscillations and for many subsequent cycles before it is handed off to an oscillation-averaged effective fluid. The stress-energy transfer described in Sec.~\ref{subsec:KG-fluid-matching} preserves the large-scale growing mode and avoids an artificial loss of matter power.

We use the representative values $m=10^{-22}\,{\rm eV}$ and $\tilde m\simeq1\,{\rm eV}$ to test, at the background and linear-perturbation levels, whether increasing the ULDM fraction can raise the value of $H_0$ required to preserve the acoustic angular scale while lowering late-time clustering. We also examine whether the accompanying CMB temperature and linear matter spectra remain regular across the KG-fluid handoff.

\begin{sloppypar}
The present analysis establishes the mechanism at the background and linear-perturbation levels and predicts a distinctive combination of a reduced sound horizon and scale-dependent suppression of structure growth. The observational consequence cannot be reduced to a single value of $\sigma_8$: cosmic shear, galaxy clustering, redshift-space distortions, and CMB lensing weight the matter power spectrum over different ranges of scale and redshift~\cite{Planck:2018vyg,KiDS:2020suj,DES:2021wwk,ACT:2023kun}. A joint multi-probe analysis provides a discriminating test of the model, since the same parameter region must simultaneously reproduce the acoustic scale and the probe-dependent pattern of late-time structure growth.
\end{sloppypar}

The remainder of this paper is organized as follows. In Section~\ref{sec:II}, we describe the self-interacting scalar model, its three dynamical regimes, and the characteristic transition scales. We present the reduction of the sound horizon and the corresponding fixed-acoustic-scale shift in $H_0$ in Section~\ref{sec:hubble-tension}. In Section~\ref{sec:perturbations}, we develop the linear perturbation evolution, the KG-to-fluid matching prescription, and the resulting CMB and matter spectra. We summarize the implications and observational prospects in Sec.~\ref{sec:Dis}.


\section{Self-interacting ultralight dark matter}
\label{sec:II} 

We adopt the metric signature $(-,+,+,+)$ throughout and consider a real self-interacting ultralight scalar field described by
\beq
S_\phi = \int d^4x\,\sqrt{-g} \left[ -\frac{1}{2}g^{\mu\nu}\partial_\mu\phi\,\partial_\nu\phi - V(\phi) \right], \qquad
V(\phi) = \frac{1}{2}m^2\phi^2 + \frac{\lambda}{4}\phi^4, \qquad \lambda>0. \label{eq:scalar}
\eeq
The full scalar field is decomposed into a spatially homogeneous background and a linear perturbation,
\beq
\phi(\mathbf{x},t) = \bar{\phi}(t) + \delta\phi(\mathbf{x},t). \label{eq:field-decomposition}
\eeq
Throughout this section, $\bar{\phi}$ denotes the homogeneous scalar background and $\delta\phi$ its linear perturbation introduced explicitly in Sec.~\ref{sec:perturbations}. For the scalar sector, we retain an overbar on background quantities, such as $\bar{\rho}_\phi$ and $\bar{p}_\phi$, in order to distinguish them explicitly from $\delta\rho_\phi$ and $\delta p_\phi$. For the standard cosmological components, however, symbols such as $\rho_{\rm cdm}$, $\rho_{\rm b}$, and $\rho_{\rm rad}$ denote homogeneous background densities unless a perturbation symbol is shown explicitly. The microscopic mass $m$ controls the wave-like behavior in the late quadratic regime, whereas the combination
\beq
\tilde m \equiv \frac{m}{\lambda^{1/4}} \label{eq:mtilde-definition}
\eeq
sets the characteristic energy scale of the quartic-to-quadratic transition. For the signature convention adopted here, the full scalar field obeys
\beq
\Box\phi - m^2\phi - \lambda\phi^3 = 0. \label{eq:full-KG-explicit}
\eeq


\subsection{Three dynamical regimes of the homogeneous field}
\label{subsec:three-regimes} 

For the spatially homogeneous background field $\bar{\phi}(t)$, the energy density and pressure are
\beq
\bar{\rho}_\phi = \frac{1}{2}\dot{\bar{\phi}}^{\,2} + V(\bar{\phi}), \qquad \bar{p}_\phi = \frac{1}{2}\dot{\bar{\phi}}^{\,2} - V(\bar{\phi}), \label{eq:rho-p-phi}
\eeq
and the instantaneous background equation of state (EOS) parameter is
\beq
\bar{w}_\phi \equiv \frac{\bar{p}_\phi}{\bar{\rho}_\phi} = \frac{\dot{\bar{\phi}}^{\,2}/2-V(\bar{\phi})}{\dot{\bar{\phi}}^{\,2}/2+V(\bar{\phi})}. \label{eq:wphi}
\eeq
The homogeneous KG equation is
\beq
\ddot{\bar{\phi}} + 3H\dot{\bar{\phi}} + m^2\bar{\phi} + \lambda\bar{\phi}^{3} = 0. \label{eq:KG-background}
\eeq
The cosmological evolution of the homogeneous field is characterized by three successive regimes, depending on the relative importance of the Hubble friction, the quartic interaction, and the quadratic mass term.


\subsubsection{Phase I: Hubble-frozen regime, $z>z_{\rm osc}$}
\label{subsubsec:2A1} 

At sufficiently early times, the expansion rate is larger than the characteristic oscillation frequency of the field,
\beq
H \gg \omega_{\rm eff}, \qquad \omega_{\rm eff}^2 \equiv V_{,\bar{\phi}\bar{\phi}} = m^2+3\lambda\bar{\phi}^{2}. \label{eq:frozen-condition}
\eeq
The KG equation is then dominated by the Hubble-friction term,
\beq
3H\dot{\bar{\phi}} + V_{,\bar{\phi}} \simeq 0, \label{eq:slow-frozen}
\eeq
so that the field evolves only slowly. In particular,
\beq
\frac{\dot{\bar{\phi}}^{\,2}}{2} \ll V(\bar{\phi}), \label{eq:frozen-kinetic}
\eeq
and Eqs.~(\ref{eq:rho-p-phi}) and (\ref{eq:wphi}) give
\beq
\bar{\rho}_\phi \simeq V(\bar{\phi}) \simeq {\rm constant}, \qquad \bar{p}_\phi \simeq - V(\bar{\phi}), \qquad \bar{w}_\phi \simeq -1. \label{eq:phase-I-w}
\eeq
Before the onset of oscillations, the scalar behaves as a vacuum-like component. Its fractional energy density is chosen to remain small in this early regime and does not disturb the standard radiation-dominated evolution.


\subsubsection{Phase II: quartic-dominated oscillations, $z_m<z<z_{\rm osc}$}
\label{sec:2A2} 

Oscillations begin when the effective oscillation frequency becomes comparable to the Hubble rate,
\beq
\omega_{\rm eff}(\phi_{\rm osc}) \sim H(z_{\rm osc}). \label{eq:oscillation-onset}
\eeq
Here and throughout the analytic phase discussion, $z_{\rm osc}$ denotes the physical onset of rapid oscillations. It is not the later numerical KG-to-EFA handoff, which is denoted by $z_{\rm handoff}$ in Sec.~\ref{subsec:KG-fluid-matching}.
When the field amplitude is sufficiently large,
\beq
\lambda\Phi^2 \gg m^2, \label{eq:quartic-condition}
\eeq
where $\Phi$ denotes the oscillation amplitude of the homogeneous scalar field $\bar\phi$, the quartic term dominates and the KG equation becomes approximately
\beq
\ddot{\bar{\phi}} + 3H\dot{\bar{\phi}} + \lambda\bar{\phi}^{3} \simeq 0. \label{eq:KG-quartic}
\eeq
During one microscopic oscillation, the expansion of the Universe is slow, $\omega_{\rm eff}\gg H$, and the Hubble-friction term can be neglected when deriving the cycle-averaged virial relation,
\beq
\left\langle \dot{\bar\phi}^{\,2}\right\rangle = \left\langle \bar\phi V_{,\bar\phi}\right\rangle . \label{eq:virial}
\eeq
This follows by multiplying the frictionless oscillation equation $\ddot{\bar\phi}+V_{,\bar\phi}\simeq0$ by $\bar\phi$ and averaging over one period. Since
\beq
\bar\phi\ddot{\bar\phi} = \frac{d}{dt}\left(\bar\phi\dot{\bar\phi}\right) - \dot{\bar\phi}^{\,2},
\eeq
the total-derivative term vanishes over a complete cycle, yielding Eq.~(\ref{eq:virial}). Equivalently, $2\langle K\rangle=\langle\bar\phi V_{,\bar\phi}\rangle$. For a monomial potential
\beq
V(\bar\phi)=A|\bar\phi|^n,
\eeq
one has $\bar\phi V_{,\bar\phi}=nV$, and thus
\beq
\left\langle\frac{\dot{\bar\phi}^{\,2}}{2}\right\rangle = \frac{n}{2}\langle V\rangle .
\eeq
The oscillation-averaged EOS is
\beq
\langle \bar{w}_\phi\rangle = \frac{n-2}{n+2}. \label{eq:w-general}
\eeq
For the quartic potential, $n=4$, giving
\beq
\langle \bar{w}_\phi\rangle = \frac{1}{3}. \label{eq:w-quartic}
\eeq
The averaged continuity equation,
\beq
\dot{\bar{\rho}}_\phi + 3H \left( 1+\langle \bar{w}_\phi\rangle \right) \bar{\rho}_\phi = 0, \label{eq:averaged-continuity}
\eeq
then yields
\beq
\bar{\rho}_\phi \propto a^{-4}, \qquad \Phi \propto a^{-1}. \label{eq:quartic-scaling}
\eeq
The oscillating scalar behaves as an additional radiation-like component during Phase II. Its effect on the pre-recombination expansion rate and on the sound-horizon integral is quantified in Secs.~\ref{subsec:sound-horizon} and \ref{subsec:mixed-background-hubble}.


\subsubsection{Phase III: quadratic-dominated oscillations, $z<z_m$}
\label{sec:2A3} 

As the Universe expands, the oscillation amplitude decreases according to Eq.~(\ref{eq:quartic-scaling}).  The transition to the quadratic regime occurs when the two terms in the potential become comparable,
\beq
\frac{1}{2}m^2\Phi_m^2 \simeq \frac{\lambda}{4}\Phi_m^4, \label{eq:quadratic-quartic-equality}
\eeq
which gives
\beq
\Phi_m \simeq \frac{\sqrt{2}\,m}{\sqrt{\lambda}}. \label{eq:phi-m}
\eeq
For $\Phi\ll\Phi_m$, the quadratic term dominates,
\beq
m^2 \gg \lambda\Phi^2, \label{eq:quadratic-condition}
\eeq
and the KG equation reduces to
\beq
\ddot{\bar{\phi}} + 3H\dot{\bar{\phi}} + m^2\bar{\phi} \simeq 0. \label{eq:KG-quadratic}
\eeq
In the rapidly oscillating limit, $m\gg H$, its leading solution can be written as
\beq
\bar{\phi}(t) \simeq \Phi(t) \cos\left(mt+\delta\right), \qquad \Phi(t) \propto a^{-3/2}. \label{eq:quadratic-solution}
\eeq
The averaged kinetic and potential energies are equal,
\beq
\left\langle \frac{\dot{\bar{\phi}}^{\,2}}{2} \right\rangle = \left\langle \frac{m^2\bar{\phi}^{2}}{2} \right\rangle, \label{eq:quadratic-virial}
\eeq
so that
\beq
\langle \bar{w}_\phi\rangle = 0 , \qquad \bar{\rho}_\phi \propto a^{-3}. \label{eq:w-quadratic}
\eeq
In this regime, the self-interacting scalar evolves as a pressureless matter component.

The complete background sequence is
\beq
\langle \bar{w}_\phi\rangle \simeq \begin{cases}
-1, & z>z_{\rm osc}, \\[1mm]
1/3, & z_m<z<z_{\rm osc}, \\[1mm]
0, & z<z_m,
\end{cases} \label{eq:three-phase-w}
\eeq
with the corresponding approximate density scaling
\beq
\bar{\rho}_\phi(a) \propto \begin{cases}
a^0, & z>z_{\rm osc}, \\[1mm]
a^{-4}, & z_m<z<z_{\rm osc}, \\[1mm]
a^{-3}, & z<z_m. \end{cases} \label{eq:three-phase-rho}
\eeq
These relations, rather than a schematic potential diagram, provide the background description used in the cosmological analysis below.


\subsection{Characteristic transition scales}
\label{subsec:transition-scales} 

We next estimate the characteristic temperatures and redshifts associated with the onset of oscillations and with the quartic-to-quadratic transition. For an adiabatically expanding thermal bath, conservation of comoving entropy gives
\beq
s(T)a^3={\rm constant}, \qquad s(T)=\frac{2\pi^2}{45}g_{*s}(T)T^3 . \label{eq:entropy-conservation}
\eeq
Defining the present-day entropy degrees of freedom by $g_{*s,0}\equiv g_{*s}(T_0)$, this gives
\beq
g_{*s}(T)T^3a^3 = g_{*s,0}T_0^3a_0^3 .
\eeq
Taking $a_0=1$ and $1+z=a^{-1}$ gives
\beq
1+z(T) = \frac{T}{T_0} \left[ \frac{g_{*s}(T)}{g_{*s,0}} \right]^{1/3}, \label{eq:z-temperature-general}
\eeq
where $T_0=2.35\times10^{-4}\,{\rm eV}$ is the present photon temperature. For the order-of-magnitude estimates below, we neglect the variation of $g_{*s}$ unless explicitly stated.


\subsubsection{Onset of quartic oscillations}
\label{subsubsec:2B1} 

In the quartic-dominated regime given in \ref{sec:2A2} ,
\beq
V(\bar{\phi})\simeq\frac{\lambda}{4}\bar{\phi}^{4}, \qquad V_{,\bar{\phi}\bar{\phi}}=3\lambda\bar{\phi}^{2}.
\eeq
The characteristic oscillation frequency is
\beq
\omega_{\rm eff} \simeq \sqrt{3\lambda}\,\Phi . \label{eq:quartic-frequency}
\eeq
The field begins to oscillate when the Hubble friction becomes comparable to this frequency,
\beq
H_{\rm osc} \simeq \kappa_{\rm osc}\sqrt{\lambda}\,\Phi_{\rm osc}, \label{eq:Hosc-condition}
\eeq
where $\kappa_{\rm osc}$ is an order-unity coefficient whose precise value depends on the convention used to define the onset.

During radiation domination, the total energy density is
\beq
\rho_r(T) = \frac{\pi^2}{30}g_*(T)T^4 . \label{eq:rho-radiation-T}
\eeq
The Friedmann equation gives
\beq
H^2(T) = \frac{8\pi G}{3}\rho_r(T) = \frac{\rho_r(T)}{3m_P^2}, \qquad m_P\equiv(8\pi G)^{-1/2}, \label{eq:friedmann-radiation}
\eeq
or
\beq
H(T) = \left( \frac{\pi^2g_*(T)}{90} \right)^{1/2} \frac{T^2}{m_P}, \label{eq:H-radiation}
\eeq
where $m_P=2.4\times10^{18}\,{\rm GeV}$ is the reduced Planck mass. Combining Eqs.~(\ref{eq:Hosc-condition}) and
(\ref{eq:H-radiation}) gives
\beq
T_{\rm osc} \simeq \left[ \frac{\kappa_{\rm osc}\sqrt{\lambda}\,  \Phi_{\rm osc}m_P} {(\pi^2g_{*,\rm osc}/90)^{1/2}} \right]^{1/2}. \label{eq:Tosc-general}
\eeq
Equivalently, using
$\bar{\rho}_{\bar{\phi},\rm osc}\simeq\lambda\Phi_{\rm osc}^4/4$,
\beq
T_{\rm osc} \simeq \left[ \frac{2\kappa_{\rm osc}m_P\sqrt{\bar{\rho}_{\bar{\phi},\rm osc}}}{(\pi^2g_{*,\rm osc}/90)^{1/2}\Phi_{\rm osc}}\right]^{1/2},\label{eq:Tosc-rho}
\eeq
up to the same order-unity onset convention.


\subsubsection{Quartic-to-quadratic transition}
\label{subsubsec:2B2} 

The quartic-phase scaling and the condition defining the quartic-to-quadratic transition were derived in Eqs.~(\ref{eq:quartic-scaling})--(\ref{eq:phi-m}). Using also the definition $\tilde m=m\lambda^{-1/4}$ from Eq.~(\ref{eq:mtilde-definition}), the two potential contributions at $\Phi=\Phi_m$ are
\beq
\frac{1}{2}m^2\Phi_m^2 = \frac{\lambda}{4}\Phi_m^4 = \frac{m^4}{\lambda} = \tilde m^4 . \label{eq:transition-components}
\eeq
The total potential energy evaluated at the equality amplitude is $2\tilde m^4$. Since the transition takes place over a finite range of oscillation amplitudes, it is convenient to write the transition density as
\beq
\bar{\rho}_{\bar{\phi},m} = C_m\tilde m^4, \label{eq:rho-transition-Cm}
\eeq
where $C_m$ is an order-unity matching coefficient. The simple instantaneous-transition convention corresponds to $C_m\simeq1$--$2$.

For the benchmark regime considered here, $z_m>z_{\rm eq}$, so the scalar is already matter-like between the transition and equality. Using the pressureless scaling in Eq.~\eqref{eq:w-quadratic} after the transition therefore gives
\beq
\bar{\rho}_{\bar{\phi},m} = \bar{\rho}_{\bar{\phi},\rm eq} \left( \frac{a_{\rm eq}}{a_m} \right)^3 . \label{eq:rho-m-aeq}
\eeq
From Eq.~(\ref{eq:entropy-conservation}),
\beq
\frac{a_{\rm eq}}{a_m} = \frac{T_m}{T_{\rm eq}} \left[ \frac{g_{*s}(T_m)}{g_{*s}(T_{\rm eq})} \right]^{1/3}, \label{eq:aeq-am-entropy}
\eeq
so that
\beq
\bar{\rho}_{\bar{\phi},m} = \bar{\rho}_{\bar{\phi},\rm eq} \left( \frac{T_m}{T_{\rm eq}} \right)^3 \frac{g_{*s}(T_m)}{g_{*s}(T_{\rm eq})}. \label{eq:rho-m-rho-eq}
\eeq
Neglecting the small change in $g_{*s}$ around equality,
\beq
T_m \simeq T_{\rm eq} \left( \frac{C_m\tilde m^4}{\bar{\rho}_{\bar{\phi},\rm eq}} \right)^{1/3}. \label{eq:Tm-general}
\eeq

For a mixed CDM+ULDM cosmology, define the present-day scalar fraction by
\beq
f_{\rm UDM} \equiv \frac{\Omega_{\phi,0}}{\Omega_{{\rm dm},0}}. \label{eq:f-udm-definition}
\eeq
Throughout this paper, $f_{\rm UDM}$ denotes the present-day fraction of the total DM density carried by the scalar. Because both the scalar and CDM scale as $a^{-3}$ after $z_m$ in the benchmark regime, this ratio is unchanged between equality and today, and hence
\beq
\bar{\rho}_{\bar{\phi},\rm eq} = f_{\rm UDM}\, \rho_{{\rm dm},\rm eq}, \label{eq:rho-phi-eq-mixed}
\eeq
where $\rho_{{\rm dm},\rm eq}$ denotes the total DM density at equality.  At matter-radiation equality,
\beq
\rho_{{\rm m},\rm eq} = \rho_{{\rm rad},\rm eq}, \qquad \rho_{{\rm rad},\rm eq} = \frac{\pi^2}{30}g_{*,\rm eq}T_{\rm eq}^4, \label{eq:equality-density}
\eeq
but $\bar{\rho}_{\bar{\phi},\rm eq}$ equals this quantity only in the limiting case in which the scalar accounts for essentially the entire matter component. For the mixed models considered here, $\bar{\rho}_{\bar{\phi},\rm eq}$ must instead be normalized to the specified present-day value of $f_{\rm UDM}$. If parameters were chosen such that $z_m<z_{\rm eq}$, the scalar would still be radiation-like at equality, and this analytic relation would require modification.

The corresponding transition redshift is
\beq
1+z_m \simeq \frac{T_m}{T_0} \left[ \frac{g_{*s}(T_m)}{g_{*s,0}} \right]^{1/3}. \label{eq:zm-general}
\eeq
The transition redshift $z_m$ depends not only on $\tilde m$, but also on the scalar fraction through $\bar{\rho}_{\bar{\phi},\rm eq}$. In particular,
\beq
1+z_m \propto \tilde m^{4/3} f_{\rm UDM}^{-1/3}, \label{eq:zm-scaling}
\eeq
when the remaining cosmological parameters are held fixed.


\subsubsection{Relation between $T_{\rm osc}$ and $T_m$}
\label{subsubsec:2B3} 

Using the quartic-phase amplitude scaling in Eq.~\eqref{eq:quartic-scaling},
\beq
\frac{\Phi_{\rm osc}}{\Phi_m} = \frac{a_m}{a_{\rm osc}} \simeq \frac{T_{\rm osc}}{T_m} \left[ \frac{g_{*s}(T_{\rm osc})}{g_{*s}(T_m)} \right]^{1/3}. \label{eq:Phi-osc-Phi-m}
\eeq
Neglecting the entropy correction for an analytic estimate and substituting Eq.~\eqref{eq:phi-m} into the onset condition gives
\beq
T_{\rm osc} \simeq \frac{\sqrt{2}\,\kappa_{\rm osc}\, m\,m_P}{(\pi^2g_{*,\rm osc}/90)^{1/2} T_m}. \label{eq:Tosc-from-Tm}
\eeq
This form makes explicit that the onset temperature depends on both the microscopic mass $m$ and the background transition temperature $T_m$.

For the representative parameter region
\beq
m\sim10^{-22}\,{\rm eV}, \qquad \tilde m\sim1\,{\rm eV}, \qquad \lambda=\left(\frac{m}{\tilde m}\right)^4 \sim10^{-88},
\eeq
the physical onset of rapid oscillations in the validated representative KG evolution, defined by $\omega_{\rm eff}/H=1$, occurs at
\beq
z_{\rm phys,osc}\simeq1.1\times10^9. \label{eq:fiducial-zphysosc}
\eeq
We do not quote a numerical onset temperature because it is not independently fixed by the final benchmark records. The precise onset depends on the adopted onset convention and on the thermal history. We do not impose a BBN or $\Delta N_{\rm eff}$ constraint on the benchmark models in the present analysis. Their detailed BBN implications require a dedicated treatment of the scalar fraction and thermal degrees of freedom and are outside the scope of the present benchmark calculation. The late-time sound-horizon modification is controlled mainly by the radiation-like phase near $z_m$ and is comparatively insensitive to the exact value of $z_{\rm phys,osc}$ provided that an extended quartic oscillatory interval is present.

\begin{figure*}[t]
\centering
\includegraphics[width=0.98\textwidth]{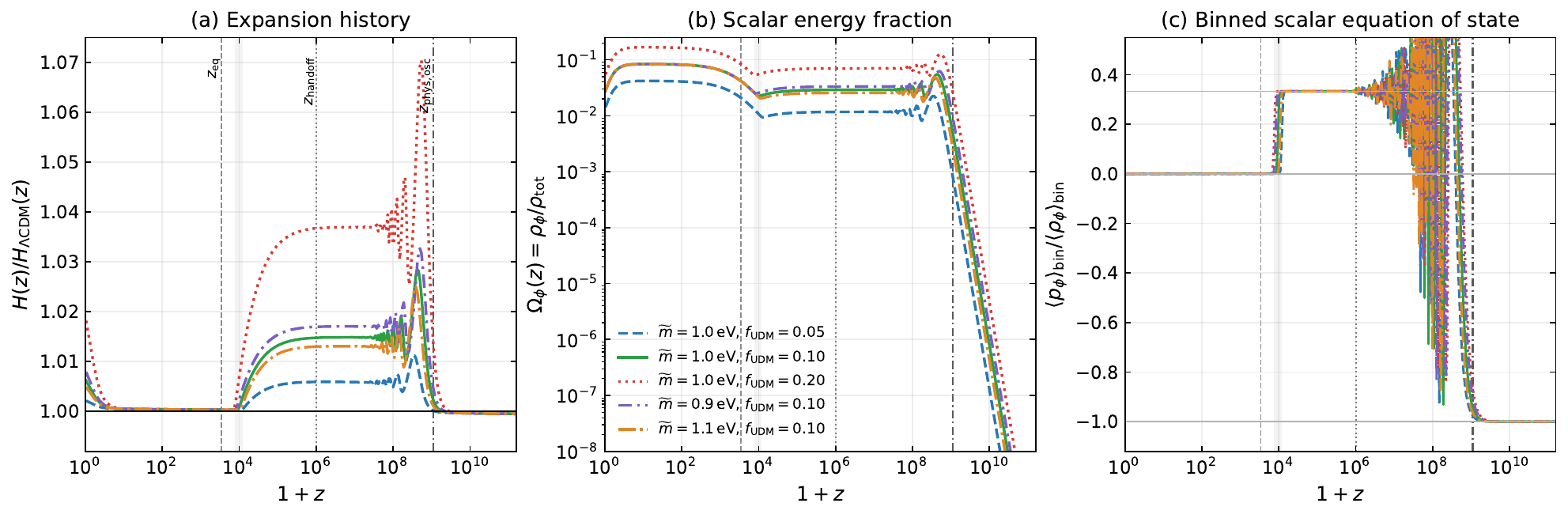}
\caption{Final background evolution for the six validated benchmarks. Panel (a) shows $H(z)/H_{\Lambda{\rm CDM}}(z)$ for the five mixed models, panel (b) shows their scalar energy fractions, and panel (c) shows the binned scalar EOS. The vertical dotted lines indicate matter-radiation equality, the numerical KG-to-EFA handoff $z_{\rm handoff}=10^6$, and the representative physical oscillation onset $z_{\rm phys,osc}\simeq1.1\times10^9$. The light shaded band marks the range $7.6\times10^3\lesssim z_m\lesssim1.21\times10^4$ spanned by the five mixed benchmarks. The $\Lambda$CDM background supplies the reference curve in panel (a).}
\label{fig:CLASS-background}
\end{figure*}

Figure~\ref{fig:CLASS-background} displays the final background evolution for the complete six-case benchmark set. Panel (a) shows the expansion-rate ratio of each mixed model to the $\Lambda$CDM reference. The excess expansion is largest for $f_{\rm UDM}=0.20$ and decreases successively through $f_{\rm UDM}=0.10$ and $0.05$ at fixed $\tilde m=1.0\,{\rm eV}$. At fixed $f_{\rm UDM}=0.10$, changing $\tilde m$ shifts the quartic-to-quadratic transition and produces the smoother, smaller separation among the $0.9$, $1.0$, and $1.1\,{\rm eV}$ histories.

Panel (b) shows the scalar fraction $\Omega_\phi(z)=\bar\rho_\phi/\rho_{\rm tot}$ for all five mixed cases. Increasing the present-day ULDM fraction raises the scalar contribution throughout the radiation-like and matter-like regimes. Varying $\tilde m$ at fixed fraction changes primarily the transition timing. Panel (c) shows the binned effective EOS: $\bar w_\phi\simeq0$ below $z_m$, $\bar w_\phi\simeq1/3$ through the quartic oscillatory regime, and $\bar w_\phi\simeq-1$ before physical oscillations begin. The vertical dotted lines indicate $z_{\rm phys,osc}\simeq1.1\times10^9$, the numerical handoff $z_{\rm handoff}=10^6$, and matter--radiation equality, while the light shaded band spans the full range of case-dependent $z_m$ values. Thus the physical oscillation onset, numerical representation switch, and physical quartic-to-quadratic transition remain separate epochs.

For a dominant or order-unity ULDM fraction, consistency with the standard matter-radiation transition requires the scalar to be approximately matter-like by equality,
\beq
z_m\gtrsim z_{\rm eq}, \label{eq:transition-before-equality}
\eeq
or equivalently $T_m\gtrsim T_{\rm eq}$. Through Eq.~(\ref{eq:Tm-general}), this condition gives the approximate scaling
\beq
\tilde m_{\min} \propto f_{\rm UDM}^{1/4} C_m^{-1/4}. \label{eq:mtilde-lower-bound-scaling}
\eeq
The familiar eV-scale lower bound is most directly applicable when the scalar constitutes most or all of the DM. For a subdominant ULDM component, the condition is relaxed because any residual radiation-like contribution near equality is weighted by $f_{\rm UDM}$. The precise bound must ultimately be obtained from the CMB and large-scale structure likelihoods rather than from the equality criterion alone.

The parameter $\tilde m$ also controls the self-interaction-supported length scale in the nonrelativistic limit. The KG equation then reduces to a Gross--Pitaevskii--Poisson or Schr\"odinger--Poisson system, depending on whether the quartic self-interaction is retained. In the repulsive self-interaction-dominated regime, the associated characteristic length is
\beq
\lambda_J = \sqrt{ \frac{3\pi\hbar^3\lambda}{4cGm^4}} = 1.2\,{\rm kpc} \left( \frac{\tilde m}{10\,{\rm eV}} \right)^{-2}. \label{eq:self-interaction-Jeans}
\eeq
This scale characterizes a self-interaction-supported Jeans scale. It should not be identified with thermal free streaming or with the scale-dependent quantum Jeans scale of a free fuzzy-DM field. Representative scalar-DM analyses associate kpc-scale cores with self-interaction strengths corresponding, in the present convention, to $\widetilde m$ values of order a few eV~\cite{Li:2013nal,Hartman:2021upg}. These estimates cannot be directly applied to our mixed CDM+ULDM scenario with $f_{\rm UDM}<1$ and are quoted only as reference values, since the actual core structure depends on the spatial distributions and combined gravitational potential of the ULDM, CDM, and baryonic components.

ULDM is not generically a thermal hot dark matter (HDM) relic. In the scenario considered here, the field is treated as a coherent condensate whose quadratic oscillations have $\langle w_\phi\rangle \simeq 0$ at the background level. Its suppression of small-scale clustering arises from gradient and effective-pressure effects, rather than from thermal velocities. Nevertheless, if the scalar constitutes all of the DM, these effects can suppress structure too strongly on small scales. The associated astrophysical bounds weaken in a mixed CDM+ULDM cosmology as $f_{\rm UDM}$ is reduced\cite{Li:2013nal,Hartman:2021upg,Diez-Tejedor:2014naa,Boudon:2022dxi}. In the present work, $\widetilde m$ and $f_{\rm UDM}$ are treated as independent phenomenological parameters, and the viable fraction must be determined by jointly testing the background expansion and the scale-dependent perturbation evolution.



\section{Sound-horizon reduction at fixed acoustic scale}
\label{sec:hubble-tension} 

The principal background effect of the self-interacting scalar field is a temporary enhancement of the pre-recombination expansion rate during its quartic, radiation-like phase. This additional energy density reduces the comoving sound horizon, while the same scalar field subsequently becomes matter-like and contributes to the present DM abundance.


\subsection{Sound horizon and acoustic angular scale}
\label{subsec:sound-horizon} 

The comoving sound horizon at photon decoupling is
\beq
r_s(z_*) = \int_0^{t_*} \frac{c_s(t)}{a(t)}\,dt = \int_{z_*}^{\infty} \frac{c_s(z)}{H(z)}\,dz , \label{eq:sound-horizon}
\eeq
where $z_*$ denotes the photon-decoupling redshift. The sound speed of the tightly coupled photon--baryon fluid is
\beq
c_s(z) = \frac{c}{\sqrt{3[1+R(z)]}}, \qquad R(z) \equiv \frac{3\rho_b(z)}{4\rho_\gamma(z)} =\frac{3\omega_b}{4\omega_\gamma(1+z)} , \label{eq:photon-baryon-sound-speed}
\eeq
where $\omega_i\equiv\Omega_i h^2$. As specified in Sec.~\ref{sec:II}, $\rho_b$ and $\rho_\gamma$ denote homogeneous background densities. The full redshift dependence of $R(z)$ is retained in the numerical calculation.

The primary CMB constrains the angular size of the sound horizon,
\beq
\theta_* = \frac{r_s(z_*)}{D_M(z_*)}, \label{eq:theta-star}
\eeq
where the comoving angular-diameter distance in a spatially flat Universe is
\beq
D_M(z_*) = c\int_0^{z_*}\frac{dz}{H(z)}. \label{eq:comoving-angular-distance}
\eeq
The quantity measured most directly by the acoustic-peak locations is $\theta_*$ rather than $H_0$ itself. A cosmological model that decreases $r_s(z_*)$ can accommodate a larger fixed-$\theta_*$ value of $H_0$, provided that $D_M(z_*)$ decreases by a corresponding fraction.

This relation should not be interpreted as a simple algebraic inverse proportionality between $r_s$ and $H_0$. The decoupling redshift, recombination history, physical matter densities, and the full expansion history all enter the determination of $r_s$ and $D_M$. We obtain the value of $H_0$ required to preserve the acoustic angular scale by solving the complete background and recombination equations and imposing the reference value of $\theta_*$ numerically. Fixing $\theta_*$ aligns the acoustic scale but does not preserve the photon-diffusion scale, the relative peak amplitudes, or the early-ISW contribution; these effects are reflected in the residual TT spectrum shown in Fig.~\ref{fig:final-tt}, and a full TT/TE/EE likelihood is beyond the present scope.


\subsection{Expansion history in the mixed CDM+ULDM scenario}
\label{subsec:mixed-background-hubble} 

We consider a mixed dark sector composed of pressureless CDM and a self-interacting ultralight scalar field. Density parameters without an explicit redshift argument denote present-day values:
\beq
\Omega_{{\rm dm},0} = \Omega_{{\rm cdm},0} + \Omega_{\phi,0}, \qquad f_{\rm UDM} \equiv \frac{\Omega_{\phi,0}}{\Omega_{{\rm dm},0}}, \label{eq:mixed-dark-sector}
\eeq
so that
\beq
\Omega_{\phi,0} = f_{\rm UDM}\Omega_{{\rm dm},0}, \qquad \Omega_{{\rm cdm},0} = (1-f_{\rm UDM})\Omega_{{\rm dm},0}. \label{eq:mixed-dark-sector-components}
\eeq
Throughout this paper, $f_{\rm UDM}$ denotes the present-day fraction of the total DM density carried by the scalar. Thus, the plotted values $0.05$, $0.10$, and $0.20$ are present-day fractions. It is not a redshift-independent ratio during the quartic phase, because $\bar{\rho}_\phi$ and $\rho_{\rm cdm}$ then obey different scaling laws.

The homogeneous scalar background follows the three dynamical regimes already summarized in Eqs.~\eqref{eq:three-phase-w} and \eqref{eq:three-phase-rho}. In particular, the scalar is radiation-like during $z_m<z<z_{\rm osc}$, with $\langle\bar w_\phi\rangle\simeq1/3$ and $\bar\rho_\phi\propto(1+z)^4$, and becomes matter-like for $z<z_m$.

The numerical calculation follows the continuous homogeneous scalar-field evolution obtained from the KG equation and its oscillation-averaged continuation, as displayed directly in Fig.~\ref{fig:CLASS-background}.

The total background expansion rate is
\beq
H^2(z) = \frac{8\pi G}{3}\left[\rho_r(z) + \rho_b(z) +  \rho_{\rm cdm}(z) + \bar{\rho}_\phi(z) +  \rho_\Lambda \right]. \label{eq:H-mixed-total}
\eeq
During the quartic regime, the scalar contributes as an additional radiation-like component, so
\beq
H_{\rm mix}(z) > H_{\Lambda{\rm CDM}}(z) \qquad (z_m<z<z_{\rm osc}) \label{eq:Hmix-larger}
\eeq
for otherwise comparable cosmological parameters. The sound-horizon integrand is then reduced,
\beq
\frac{c_s(z)}{H_{\rm mix}(z)} < \frac{c_s(z)}{H_{\Lambda{\rm CDM}}(z)}, \label{eq:sound-integrand-reduction}
\eeq
which implies
\beq
r_s^{\rm mix}(z_*) < r_s^{\Lambda{\rm CDM}}(z_*). \label{eq:sound-horizon-reduction}
\eeq

The modification is most efficient when the radiation-like interval overlaps the range of redshifts that contributes appreciably to the sound-horizon integral. This occurs when the quartic-to-quadratic transition is not far above recombination and matter--radiation equality. The result is less sensitive to the exact value of $z_{\rm osc}$ once $z_{\rm osc}\gg z_m$, because the homogeneous scalar density is subdominant during the early frozen phase and the sound-horizon integral is weighted more strongly toward its lower-redshift end.


\subsection{Numerical determination at fixed acoustic scale}
\label{subsec:fixed-theta-H0}

For each choice of
\beq
\left( m_{\rm ULDM}, \tilde m, f_{\rm UDM} \right),
\eeq
the CLASS calculation determines the homogeneous scalar-field evolution, the total expansion history, the recombination redshift, the sound horizon, and the comoving distance self-consistently. The Hubble constant is then adjusted until
\beq
\theta_*^{\rm mix} = \theta_*^{\Lambda{\rm CDM}}. \label{eq:fixed-theta-shooting}
\eeq
This procedure does not assume that $H(z_*)$, $z_*$, or the physical density parameters are model independent.

For the benchmark microscopic mass
\beq
m_{\rm ULDM} = 10^{-22}\,{\rm eV}, \label{eq:benchmark-ULDM-mass}
\eeq
and the six validated fixed-acoustic-scale solutions are collected in Table~\ref{tab:final-six-benchmarks}. The realized ULDM densities agree with their targets to the accuracy documented in the numerical validation record.

\begin{table*}[t]
\caption{Final validated benchmark backgrounds and linear clustering amplitudes. Distances are in Mpc and $H_0$ is in ${\rm km\,s^{-1}\,Mpc^{-1}}$. A dash denotes a quantity not applicable to the $\Lambda$CDM reference.}
\label{tab:final-six-benchmarks}
\centering
\scriptsize
\resizebox{\textwidth}{!}{%
\begin{tabular}{c|c|c|c|c|c|c|c|c|c}
Model & $\tilde m\,[{\rm eV}]$ & $f_{\rm UDM}$ & $H_0$ & $z_m$ & $z_{\rm eq}$ & $r_s(z_{\rm rec})$ & $r_{\rm drag}$ & $100\theta_s$ & $\sigma_8$ \\
\hline
$\Lambda$CDM & -- & 0 & 67.8316 & -- & 3402.875 & 144.5479 & 147.1142 & 1.04178254 & 0.824024 \\
MT100\_F005 & 1.0 & 0.05 & 67.9700 & 12070.015 & 3402.873 & 144.4907 & 147.0569 & 1.04178302 & 0.780514 \\
MT100\_F010 & 1.0 & 0.10 & 68.2541 & 9579.771 & 3402.875 & 144.3734 & 146.9396 & 1.04178305 & 0.727754 \\
MT100\_F020 & 1.0 & 0.20 & 69.1022 & 7603.263 & 3402.864 & 144.0259 & 146.5922 & 1.04178302 & 0.619318 \\
MT090\_F010 & 0.9 & 0.10 & 68.3780 & 8324.119 & 3402.869 & 144.3225 & 146.8887 & 1.04178301 & 0.721245 \\
MT110\_F010 & 1.1 & 0.10 & 68.1655 & 10878.043 & 3402.871 & 144.4099 & 146.9761 & 1.04178257 & 0.733621
\end{tabular}}
\end{table*}

For this benchmark sequence,
\beq
f_{\rm UDM}\uparrow \quad\Longrightarrow\quad H_0\uparrow, \qquad \sigma_8\downarrow. \label{eq:correlated-H0-sigma8}
\eeq
The increase of $H_0$ originates from the reduction of the sound horizon, whereas the change of $\sigma_8$ is a perturbative effect and is discussed separately in Sec.~\ref{sec:perturbations}.

At fixed $f_{\rm UDM}=0.10$, the values
\beq
\tilde m = 0.9,\ 1.0,\ 1.1\ {\rm eV}
\eeq
give
\beq
H_0 = 68.3780,\ 68.2541,\ 68.1655\ {\rm km\,s^{-1}\,Mpc^{-1}}, \label{eq:H0-three-mtilde}
\eeq
respectively. In the same ordering,
\beq
\begin{aligned}
r_s(z_{\rm rec})&=(144.3225,144.3734,144.4099)\,{\rm Mpc},\\
\sigma_8&=(0.721245,0.727754,0.733621).
\end{aligned}
\eeq
Thus, increasing $\tilde m$ smoothly lowers $H_0$, increases the sound horizon and clustering amplitude, and weakens matter-power suppression. The magnitude of the response is controlled primarily by the scalar fraction, while the dependence on $\tilde m$ is milder over this interval. This weaker dependence is a local property of the explored $0.9$--$1.1\,{\rm eV}$ interval and should not be extrapolated to a broader range of $\tilde m$.

The corresponding CMB temperature and matter-power spectra are presented and interpreted in Sec.~\ref{sec:perturbations}. Figure~\ref{fig:final-tt} displays both benchmark scans for the unlensed TT spectrum, and Fig.~\ref{fig:final-pk} displays the same two scans for the linear matter power spectrum.

A free ultralight scalar with $\lambda=0$ can produce the familiar scale-dependent growth suppression associated with wave pressure, but it does not possess the quartic radiation-like phase required for the sound-horizon reduction described above. The repulsive self-interaction is essential to the background mechanism of the present model.


\section{Perturbations, KG-fluid matching, and linear spectra}
\label{sec:perturbations} 

The background mechanism described above becomes observationally meaningful only when perturbations of the self-interacting ULDM component are evolved consistently. The homogeneous scalar $\bar\phi$ and perturbation $\delta\phi$ are first evolved with the exact KG system. At the configured numerical handoff, long after rapid oscillations have physically begun, the calculation changes representation to an oscillation-averaged effective fluid. Accurate Boltzmann calculations of ultralight scalar fields require particular care in this replacement because an inconsistent handoff can generate spurious features in the CMB and matter transfer functions~\cite{Urena-Lopez:2015gur,Cookmeyer:2019rna,Passaglia:2022bcr,Liu:2024yne,Moss:2025ymr}.


\subsection{Effective fluid and the large-scale limit}
\label{subsec:effective-fluid} 

For a rapidly oscillating scalar field, the effective fluid has a scale-dependent rest-frame sound speed. In natural units, $c=\hbar=1$, the free-field gradient term is the dimensionless rest-frame sound speed associated with the physical wavenumber $k/a$. In the quadratic-dominated regime, it has the leading form
\beq
c_{s,\phi}^{2}(k,a) \simeq \frac{k^2}{4m^2a^2}. \label{eq:cs2-quadratic}
\eeq
Appendix~\ref{app:si-units} gives the corresponding SI restoration. For the potential considered here, Cembranos et al.~\cite{Cembranos:2018ulm} find that the leading result in the quadratic-dominated, weakly anharmonic regime is
\beq
c_{s,\phi}^{2}(k,a) \simeq \frac{k^2}{4m^2a^2} + \frac{3}{4}\frac{\lambda\bar{\rho}_\phi}{m^4}, \label{eq:cs2-self-leading}
\eeq
Motivated by this leading result and by the limiting behavior discussed in the same reference, we use the effective interpolation
\beq
c_{s,\phi}^{2}(k,a) = \frac{\displaystyle \frac{k^2}{4m^2a^2} + \frac{3}{4}\frac{\lambda\bar{\rho}_\phi}{m^4}}{\displaystyle 1 + \frac{k^2}{4m^2a^2} +
\frac{9}{4}\frac{\lambda\bar{\rho}_\phi}{m^4}}. \label{eq:cs2-self}
\eeq
Equation~\eqref{eq:cs2-self} is a physically motivated interpolation, not an exact closure through the quartic-to-quadratic transition. Its relevant limits are
\beq
\begin{aligned}
\frac{\lambda\bar{\rho}_\phi}{m^4}\ll1,\quad \frac{k}{ma}\ll1 &\quad\Longrightarrow\quad c_{s,\phi}^{2}\rightarrow0,\\
\frac{\lambda\bar{\rho}_\phi}{m^4}\gg1,\quad k\rightarrow0 &\quad\Longrightarrow\quad c_{s,\phi}^{2}\rightarrow\frac13.
\end{aligned}
\label{eq:large-scale-fluid-limit}
\eeq
The pressureless low-$k$ limit applies only after the field is well inside the quadratic, self-interaction-negligible regime; it does not apply during quartic-dominated oscillations. In the late regime and for $k\ll k_J(a)$, the ULDM growing mode approaches the pressureless mode. The transfer function is then CDM-like on sufficiently large spatial scales, whereas physical suppression develops only at and below the effective Jeans scale~\cite{Hu:2000ke,Marsh:2010wq,Hlozek:2014lca,Passaglia:2022bcr}.

In the low-$k$ limit, $c_{s,\phi}^2\rightarrow0$ means that the ULDM perturbation within a given cosmology approaches the pressureless growing mode. It does not require the total-matter ratio $P_{\rm mix}(k)/P_{\Lambda{\rm CDM}}(k)$ to approach exactly unity, because the two fixed-$\theta_*$ cosmologies have different background parameters. Nevertheless, the KG-fluid replacement must not introduce a discontinuous or switch-localized suppression of the long-wavelength mode.


\subsection{Stress-energy matching at the KG-fluid boundary}
\label{subsec:KG-fluid-matching} 

In our implementation, the exact background and perturbed KG systems are evolved through the physical oscillation onset at $z_{\rm phys,osc}\simeq1.1\times10^9$. The representation changes only at the configured numerical handoff
\beq
z_{\rm handoff}=z_{\rm switch}=10^6,
\eeq
when the scalar is already deep in the rapid-oscillation regime and many cycles have been resolved. The handoff redshift is therefore not an estimate of the physical onset redshift.

At the handoff, the effective-fluid state is initialized from causal oscillation averages of the KG stress-energy. Schematically,
\beq
\left(\bar\rho_\phi,\bar p_\phi,\delta\rho_\phi,q_\phi\right)_{\rm EFA}
=
\left\langle\bar\rho_\phi,\bar p_\phi,\delta\rho_\phi,q_\phi\right\rangle_{\rm causal,KG},
\qquad
q_\phi\equiv(\bar\rho_\phi+\bar p_\phi)\theta_\phi .
\label{eq:matching}
\eeq
This transfers the scalar energy and momentum sources entering the Einstein equations without selecting a special microscopic oscillation phase. All non-UDM perturbation and metric variables that do not change representation are transferred continuously. The post-handoff pressure perturbation is advanced with the oscillation-averaged effective-fluid closure
\beq
\delta p_{\phi,{\rm eff}}^{({\rm rf})}=c_{s,\phi}^{2}(k,a)\,\delta\rho_{\phi,{\rm eff}}^{({\rm rf})},
\label{eq:rest-frame-pressure-closure}
\eeq
where $c_{s,\phi}^{2}$ is given by Eq.~\eqref{eq:cs2-self}. This successful-path prescription follows the stress-energy criteria developed for effective treatments of rapidly oscillating scalar fields~\cite{Cookmeyer:2019rna,Passaglia:2022bcr,Liu:2024yne,Moss:2025ymr}.

The numerical representation is activated smoothly with width $\Delta_{\rm act}=1.25\times10^{-3}$. Representative activation-width regression tests show convergence of the background and perturbation observables, and the final benchmark spectra and transfer functions remain regular over the validated wavenumber range. These tests validate the numerical handoff without identifying it with the much earlier physical onset of oscillations.


\subsection{Mixed CDM+ULDM benchmark}
\label{subsec:mixed-benchmark}

We use the mixed dark-sector parametrization and benchmark sequence defined in Secs.~\ref{subsec:mixed-background-hubble} and \ref{subsec:fixed-theta-H0}. In particular, Eq.~\eqref{eq:mixed-dark-sector} defines $f_{\rm UDM}$, Eq.~\eqref{eq:fixed-theta-shooting} specifies the fixed-$\theta_*$ procedure, and Table~\ref{tab:final-six-benchmarks} collects the final validated outputs. The quoted $\sigma_8$ values are linear-theory clustering amplitudes and are not weak-lensing determinations of $S_8$.

\begin{figure*}[t]
\centering
\includegraphics[width=0.96\textwidth]{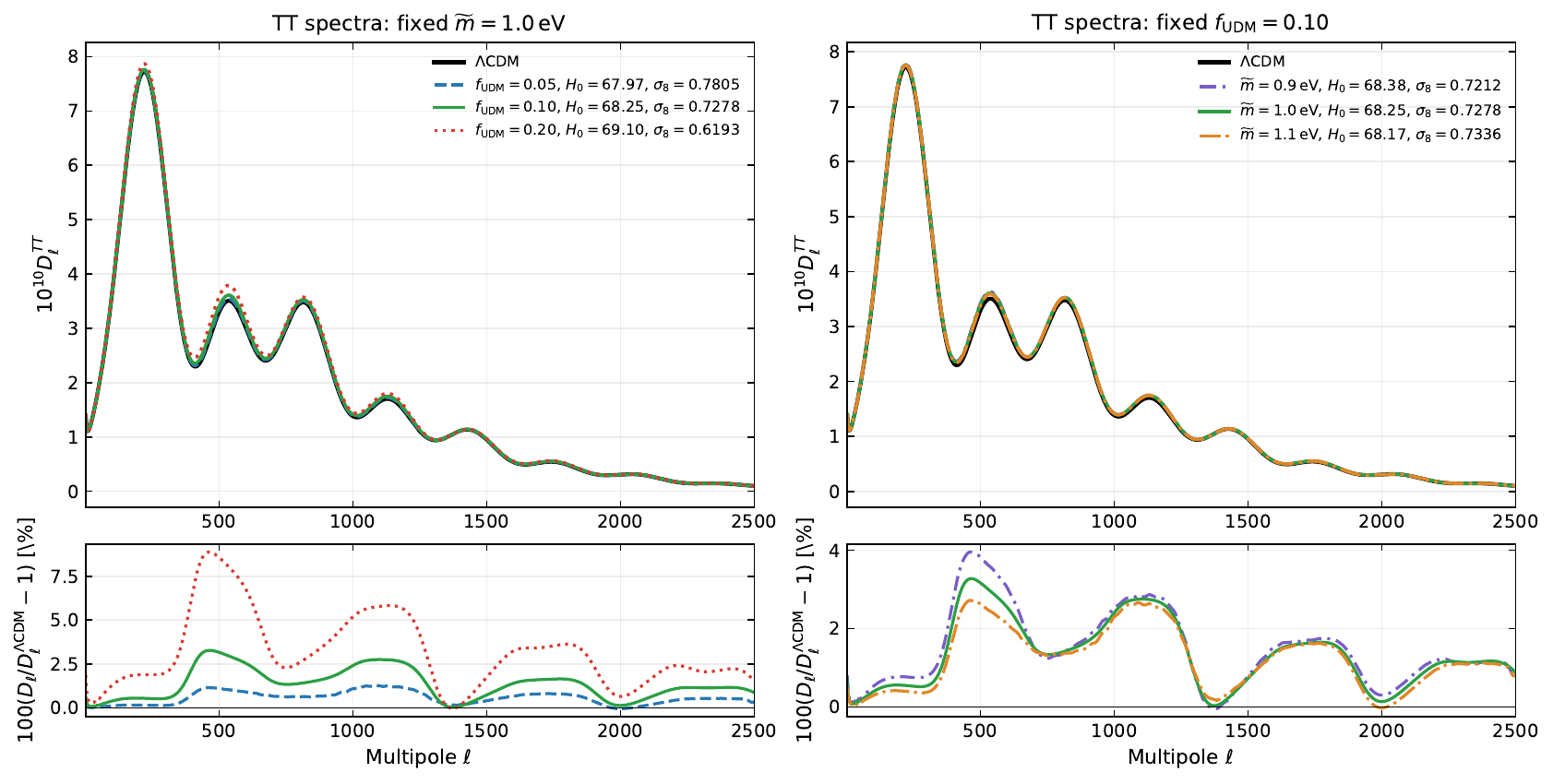}
\caption{Final unlensed CMB TT spectra for the six validated benchmarks. The left column varies $f_{\rm UDM}=0.05$, $0.10$, and $0.20$ at fixed $\tilde m=1.0\,{\rm eV}$; the right column varies $\tilde m=0.9$, $1.0$, and $1.1\,{\rm eV}$ at fixed $f_{\rm UDM}=0.10$. The upper panels show $D_\ell^{TT}$ and the lower panels show the percentage residual relative to the same $\Lambda$CDM reference. The maximum absolute residuals are $(1.28,3.28,8.87)\%$ in the fraction scan and $(3.96,3.28,2.72)\%$ in the $\tilde m$ scan.}
\label{fig:final-tt}
\end{figure*}

Figure~\ref{fig:final-tt} compares the final unlensed CMB temperature spectra and their residuals relative to $\Lambda$CDM. The left column holds $\tilde m=1.0\,{\rm eV}$ fixed and increases $f_{\rm UDM}=0.05\to0.10\to0.20$; the right column holds $f_{\rm UDM}=0.10$ fixed and increases $\tilde m=0.9\to1.0\to1.1\,{\rm eV}$. The acoustic features remain aligned by the fixed-angle construction, while their amplitudes change smoothly. In the fraction scan, the maximum absolute residual increases from $1.28\%$ to $3.28\%$ and $8.87\%$. In the $\tilde m$ scan the corresponding maxima decrease smoothly from $3.96\%$ to $3.28\%$ and $2.72\%$. Thus, increasing the fraction strengthens the TT response, whereas moving the transition earlier by increasing $\tilde m$ weakens it over the tested interval.

\begin{figure*}[t]
\centering
\includegraphics[width=0.96\textwidth]{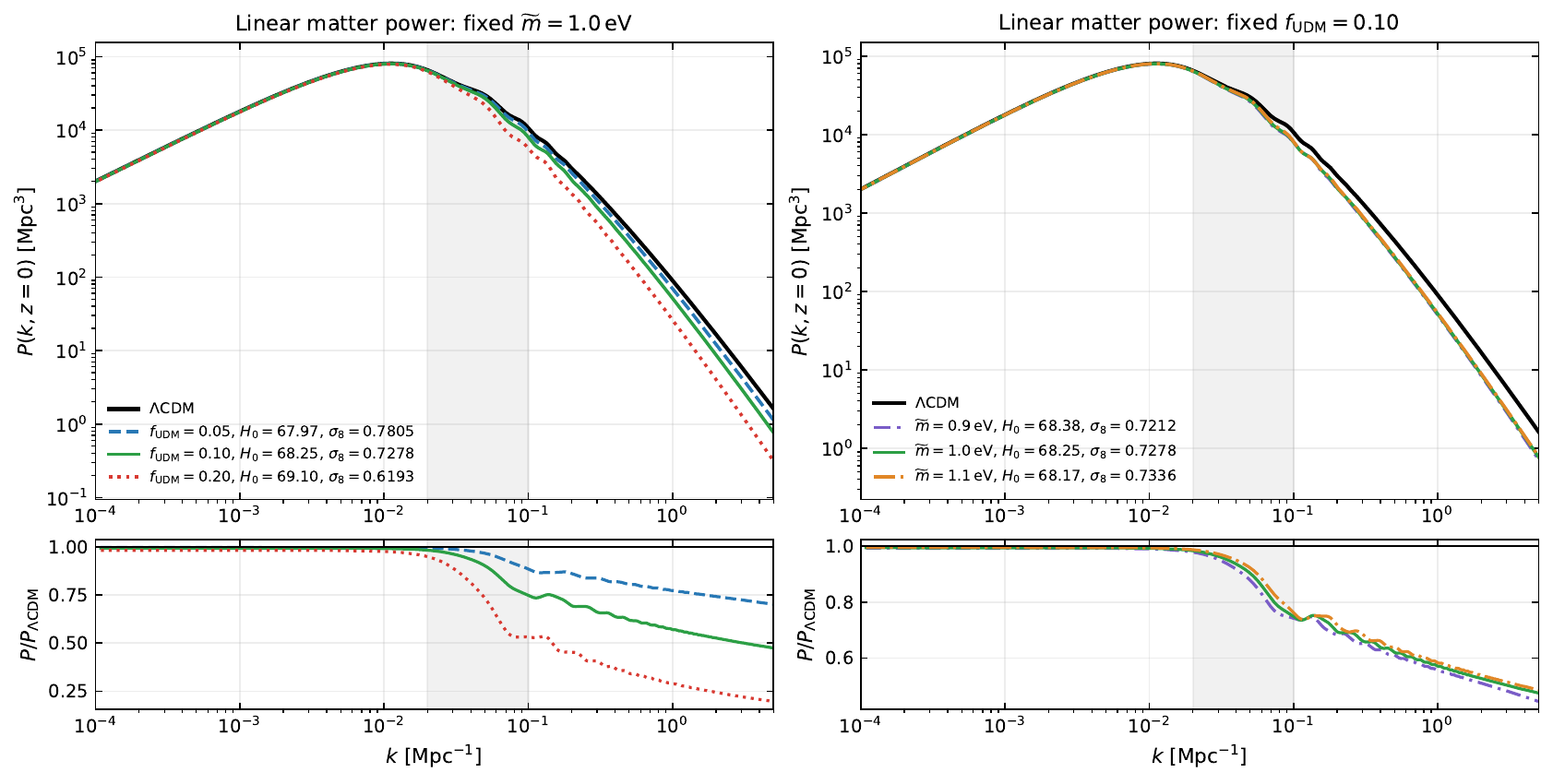}
\caption{Final linear matter power spectra at $z=0$ for the six validated benchmarks. The left column varies $f_{\rm UDM}$ at fixed $\tilde m=1.0\,{\rm eV}$, and the right column varies $\tilde m$ at fixed $f_{\rm UDM}=0.10$. The upper panels show $P(k)$ in physical units and the lower panels show $P/P_{\Lambda{\rm CDM}}$. Increasing $f_{\rm UDM}$ strengthens the scale-dependent suppression and lowers $\sigma_8$; increasing $\tilde m$ over $0.9$--$1.1\,{\rm eV}$ weakens the suppression and raises $\sigma_8$ smoothly. The shaded interval is $0.02<k<0.1\,{\rm Mpc}^{-1}$.}
\label{fig:final-pk}
\end{figure*}

\begin{sloppypar}
Figure~\ref{fig:final-pk} presents the corresponding linear matter spectra at $z=0$. At fixed $\tilde m=1.0\,{\rm eV}$, increasing the scalar fraction lowers $\sigma_8=(0.780514,0.727754,0.619318)$ and strengthens the suppression continuously. At $k=0.1\,{\rm Mpc}^{-1}$ the ratios $P/P_{\Lambda{\rm CDM}}$ are $(0.885,0.748,0.532)$ for $f_{\rm UDM}=(0.05,0.10,0.20)$, and at $k=1\,{\rm Mpc}^{-1}$ they are $(0.774,0.571,0.289)$. The $f_{\rm UDM}=0.20$ benchmark therefore exhibits strong suppression.

At fixed $f_{\rm UDM}=0.10$, increasing $\tilde m=(0.9,1.0,1.1)\,{\rm eV}$ raises $\sigma_8=(0.721245,0.727754,0.733621)$ and weakens the suppression smoothly. In the same ordering, $P/P_{\Lambda{\rm CDM}}=(0.742,0.748,0.762)$ at $k=0.1\,{\rm Mpc}^{-1}$ and $(0.556,0.571,0.583)$ at $k=1\,{\rm Mpc}^{-1}$. The model curves need not approach the reference exactly at low $k$, because the six cosmologies have distinct fixed-angle backgrounds. The shaded interval in Fig.~\ref{fig:final-pk} marks $0.02<k<0.1\,{\rm Mpc}^{-1}$ in physical CLASS wavenumber units.
\end{sloppypar}

\subsection{Interpretation and scope}
\label{subsec:perturbation-interpretation}

The numerical result realizes the two required tendencies within one dark sector. The quartic radiation-like phase increases the pre-recombination expansion rate, reduces the sound horizon, and raises the value of $H_0$ required to preserve the measured acoustic angle. The late-time effective pressure of the ULDM component suppresses growth below its characteristic scale and lowers the linear clustering amplitude, while the remaining CDM component preserves large-scale clustering. This division of roles is the principal advantage of the mixed scenario.

The linear spectra provide two complementary consistency tests. First, the large-scale matter mode and the validated transfer diagnostics remain regular across the KG-fluid handoff; no zero or handoff-localized feature is present over the checked range. Second, the modification of $P(k)$ is scale dependent, as expected for a component with a finite effective sound speed. The observable consequences of this scale dependence must ultimately be tested with the distinct redshift and scale kernels of CMB lensing, galaxy clustering, redshift-space distortions, and weak lensing.


\section{Discussion and conclusions}
\label{sec:Dis} 

We have shown that a mixed CDM+self-interacting-ULDM cosmology realizes two complementary shifts through successive regimes of one scalar field. Its quartic-dominated phase reduces the sound horizon, while its later quadratic phase contributes to dark matter and suppresses growth below the effective Jeans scale. The background-plus-linear-perturbation calculation demonstrates that the fixed-$\theta_*$ value of $H_0$ can increase while the late-time clustering amplitude decreases, without introducing a separate early-dark-energy field.

The perturbation treatment is essential to this conclusion. At the numerical KG-to-EFA handoff, the causal oscillation-averaged scalar stress-energy initializes the effective fluid and unchanged non-UDM variables are transferred continuously. With this prescription and smooth activation, the large-scale growing mode remains regular and the matter transfer function contains no handoff-localized irregularity. The remaining change in $P(k)$ is smooth and scale dependent, as expected from the physical effective pressure of the ULDM component.
 
For $m_{\rm ULDM}=10^{-22}\,{\rm eV}$ and $\tilde m=1.0\,{\rm eV}$, increasing the present-day ULDM fraction from $f_{\rm UDM}=0.05$ to $0.20$ raises the fixed-acoustic-scale value of $H_0$ from $67.9700$ to $69.1022\,{\rm km\,s^{-1}\,Mpc^{-1}}$, reduces $r_s(z_{\rm rec})$ from $144.4907$ to $144.0259\,{\rm Mpc}$, and lowers the linear clustering amplitude from $\sigma_8=0.780514$ to $0.619318$. Over this range, the TT residual remains smooth but reaches approximately $8.9\%$ in the strongest benchmark, and the matter spectrum develops strong scale-dependent suppression. At fixed $f_{\rm UDM}=0.10$, increasing $\tilde m$ from $0.9$ to $1.1\,{\rm eV}$ smoothly lowers $H_0$ from $68.3780$ to $68.1655\,{\rm km\,s^{-1}\,Mpc^{-1}}$, raises $r_s(z_{\rm rec})$ from $144.3225$ to $144.4099\,{\rm Mpc}$, raises $\sigma_8$ from $0.721245$ to $0.733621$, and weakens the matter-power suppression.

The three model parameters play distinct roles. The microscopic mass sets the wave-pressure scale, $\tilde m=m/\lambda^{1/4}$ controls the quartic-to-quadratic transition epoch, and $f_{\rm UDM}$ determines how strongly the scalar affects the total dark sector. The mixed CDM+ULDM construction is important because the CDM component preserves efficient large-scale clustering, while the ULDM fraction controls the strength of both the background and perturbative modifications.

The present results should not be interpreted as a statistical resolution of either tension. In particular, a decrease in $\sigma_8$ is not by itself a measurement of the quantity inferred from weak-lensing surveys. Cosmic shear, galaxy clustering, redshift-space distortions, and CMB lensing probe different redshift and scale kernels and require the corresponding nonlinear and nuisance-parameter modeling. A joint multi-probe likelihood analysis is therefore required to determine whether the same parameter region can simultaneously fit the acoustic scale and the observed pattern of late-time structure growth.

In summary, the main result is the identification and numerical realization of a single self-interacting-ULDM mechanism that links an early reduction of the sound horizon to a later suppression of structure growth. The CLASS implementation shows that this link survives a stress-energy-consistent treatment of linear perturbations and remains compatible with a regular large-scale growing mode. The validated benchmarks exhibit simultaneous background and growth shifts in the directions relevant to the Hubble and structure-growth tensions; their quantitative observational viability remains to be established by a full multi-probe likelihood analysis.

\appendix
\section{Restoring SI units}
\label{app:si-units}

The analytic equations in the main text use the conventional relativistic and particle-physics convention in which $c=\hbar=k_{\rm B}=1$ unless an explicit dimensional constant is displayed.  The following rules restore SI units when comparing those equations with measured quantities or with the inverse-length variables used by CLASS.  We distinguish the particle mass, rest energy, Compton angular frequency, and Compton wavenumber as
\beq
m_{\rm phys}\ [{\rm kg}],\qquad
E_m\equiv m_{\rm phys}c^2\ [{\rm J\ or\ eV}],\qquad
\omega_m\equiv\frac{E_m}{\hbar}\ [{\rm s}^{-1}],\qquad
k_m\equiv\frac{E_m}{\hbar c}\ [{\rm m}^{-1}],
\label{eq:si-mass-scales}
\eeq
and define $\widetilde E_m\equiv E_m/\lambda^{1/4}$.  Thus, a value such as $m=10^{-22}\,{\rm eV}$ in the main text denotes the rest energy $E_m=m_{\rm phys}c^2$, not a mass in kilograms.  The main-text notation $m$ and $\tilde m$ is retained everywhere outside this appendix.

For the field-dependent entries below, $\phi_{\rm SI}$ denotes the canonically normalized SI field, with $[\phi_{\rm SI}]=({\rm J}/{\rm m})^{1/2}$, and $U$ is an energy density in ${\rm J\,m^{-3}}$.  We write $\varepsilon$ for an SI energy density and $\rho_{\rm mass}=\varepsilon/c^2$ for the corresponding mass density.

\clearpage
\begin{table}[p]
\centering
\scriptsize
\setlength{\tabcolsep}{3pt}
\renewcommand{\arraystretch}{1.18}
\begin{tabular}{@{}p{0.20\textwidth}p{0.25\textwidth}p{0.48\textwidth}@{}}
\hline\hline
Quantity or term & Natural-unit form used in the text & SI restoration \\
\hline
Mass value, Eq.~(\ref{eq:benchmark-ULDM-mass})
& $m=10^{-22}\,{\rm eV}$
& $E_m=m_{\rm phys}c^2=10^{-22}\,{\rm eV}$; use $m_{\rm phys}=E_m/c^2$ only when a mass in kg is required. \\
Compton scales associated with Eqs.~(\ref{eq:intro-potential}) and (\ref{eq:KG-background})
& $m$
& $\omega_m=E_m/\hbar$ and $k_m=E_m/(\hbar c)$; frequency and inverse length must not be interchanged. \\
Quadratic potential, Eq.~(\ref{eq:scalar})
& $\frac12m^2\phi^2$
& $\frac12[E_m/(\hbar c)]^2\phi_{\rm SI}^2$. \\
Quartic potential, Eq.~(\ref{eq:scalar})
& $\frac{\lambda}{4}\phi^4$
& $\lambda\phi_{\rm SI}^4/(4\hbar c)$ for dimensionless $\lambda$. \\
Homogeneous KG mass term, Eq.~(\ref{eq:KG-background})
& $m^2\bar\phi$
& $(E_m/\hbar)^2\bar\phi_{\rm SI}$ in the cosmic-time equation; the quartic force is $(\lambda c/\hbar)\bar\phi_{\rm SI}^3$. \\
Scalar density and pressure, Eq.~(\ref{eq:rho-p-phi})
& $\bar\rho_\phi=\dot{\bar\phi}^{\,2}/2+V$, $\bar p_\phi=\dot{\bar\phi}^{\,2}/2-V$
& $\bar\varepsilon_\phi=\dot{\bar\phi}_{\rm SI}^{\,2}/(2c^2)+U$ and $\bar p_\phi=\dot{\bar\phi}_{\rm SI}^{\,2}/(2c^2)-U$. \\
Friedmann equation, Eq.~(\ref{eq:H-mixed-total})
& $H^2=(8\pi G/3)\sum_i\rho_i$
& $H_{\rm SI}^2=(8\pi G/3c^2)\sum_i\varepsilon_i$ when $\varepsilon_i$ is in ${\rm J\,m^{-3}}$; equivalently, $H_{\rm SI}^2=(8\pi G/3)\sum_i\rho_{{\rm mass},i}$. \\
Transition density, Eq.~(\ref{eq:rho-transition-Cm})
& $\bar\rho_{\phi,m}=C_m\tilde m^4$
& $\bar\varepsilon_{\phi,m}=C_m\widetilde E_m^{\,4}/(\hbar c)^3$. \\
Radiation-era rate, Eq.~(\ref{eq:H-radiation})
& $H=[\pi^2g_*/90]^{1/2}T^2/m_P$
& $H_{\rm SI}=[\pi^2g_*/90]^{1/2}(k_{\rm B}T)^2/(\hbar\bar E_{\rm P})$, where $\bar E_{\rm P}=\sqrt{\hbar c^5/(8\pi G)}$.  A temperature quoted in eV means the thermal energy $k_{\rm B}T$. \\
Gradient sound-speed term, Eqs.~(\ref{eq:cs2-quadratic}) and (\ref{eq:cs2-self})
& $k^2/(4m^2a^2)$
& $\hbar^2c^2k_{\rm com,SI}^2/(4a^2E_m^2)$ in the dimensionless ratio $\widehat c_{s,\phi}^{\,2}\equiv c_{s,\phi}^2/c^2$. \\
Self-interaction sound-speed term, Eq.~(\ref{eq:cs2-self})
& $\lambda\bar\rho_\phi/m^4$
& $\lambda(\hbar c)^3\bar\varepsilon_\phi/E_m^4$. \\
Jeans/core scale, Eq.~(\ref{eq:self-interaction-Jeans})
& $\lambda_J\propto\sqrt{\lambda/m^4}$
& $\lambda_J=\sqrt{3\pi\hbar^3\lambda/(4cGm_{\rm phys}^4)}=\sqrt{3\pi\hbar^3c^7/(4G\widetilde E_m^{\,4})}$. \\
Hubble rate at the CLASS boundary
& $H$ or $\mathcal H=aH$
& $H_{\rm CLASS}[{\rm Mpc}^{-1}]=(H_{\rm SI}/c)(1\,{\rm Mpc})$ and $\mathcal H_{\rm CLASS}=aH_{\rm CLASS}$. \\
Physical/comoving wavenumber
& $k/a$ and $k$
& $k_{\rm com,SI}=a k_{\rm phys,SI}$; $k_{\rm CLASS}[{\rm Mpc}^{-1}]=k_{\rm com,SI}[{\rm m}^{-1}](1\,{\rm Mpc})$. \\
Energy density at the CLASS boundary
& $(8\pi G/3)\rho$
& $[8\pi G\varepsilon/(3c^4)](1\,{\rm Mpc})^2$ in the corresponding inverse-Mpc-squared Friedmann contribution. \\
\hline\hline
\end{tabular}
\caption{Compact restoration of SI units for the principal quantities in the main text. The wavenumber $k$ in the perturbation equations is comoving.}
\label{tab:si-restoration}
\end{table}

In particular, the two dimensionless combinations entering the effective-fluid closure are formed only after all inputs have been expressed in one unit system:
\beq
\frac{k}{ma}\ \longrightarrow\ \frac{\hbar c\,k_{\rm com,SI}}{aE_m},
\qquad
\frac{\lambda\bar\rho_\phi}{m^4}\ \longrightarrow\
\frac{\lambda(\hbar c)^3\bar\varepsilon_\phi}{E_m^4}.
\label{eq:si-fluid-combinations}
\eeq
Likewise, CLASS stores time and distance variables as lengths with $c=1$; therefore a rate in ${\rm s}^{-1}$ is divided by $c$ before conversion to ${\rm Mpc}^{-1}$.  These rules supply a conversion map for the existing equations and do not introduce a second dynamical formulation.

\section*{Data, software, and code availability}

The UDMCLASS source modification and reproducibility package used for the
calculations in this work is archived as version v1.0.0 on Zenodo at
\href{https://doi.org/10.5281/zenodo.22209422}{doi:10.5281/zenodo.22209422}~\cite{Lee:2026udmclass}.
The corresponding GitHub repository is available at
\par\noindent\url{https://github.com/skylee2/UDMCLASS-patch}.
\par\noindent The archived package contains
the certified patch against official CLASS v3.3.4, six frozen benchmark
inputs, compact reference products, and scripts for reproducing the benchmark
quantities and figures~1--3. The full upstream CLASS source tree is not
redistributed and must be obtained separately as documented in the archive.

\acknowledgments

SL is supported by the Basic Science Research Program through the National Research Foundation of Korea (NRF), funded by the Ministry of Science and ICT under the Grant No. NRF-2022R1A2C1005050. 

\paragraph{Declaration of generative AI and AI-assisted technologies.}
During the preparation of this work, the authors used ChatGPT (OpenAI) to improve the clarity and readability of the manuscript and to assist with consistency checks of the numerical implementation. The scientific ideas, equations, numerical methods, code development, and validation were carried out by the authors. All AI-assisted suggestions were critically reviewed and independently verified by the authors, who take full responsibility for the content, results, and conclusions of this work.


\begin{thebibliography}{99}
\expandafter\ifx\csname natexlab\endcsname\relax\def\natexlab#1{#1}\fi
\expandafter\ifx\csname bibnamefont\endcsname\relax
  \def\bibnamefont#1{#1}\fi
\expandafter\ifx\csname bibfnamefont\endcsname\relax
  \def\bibfnamefont#1{#1}\fi
\expandafter\ifx\csname citenamefont\endcsname\relax
  \def\citenamefont#1{#1}\fi
\expandafter\ifx\csname url\endcsname\relax
  \def\url#1{\texttt{#1}}\fi
\expandafter\ifx\csname urlprefix\endcsname\relax\def\urlprefix{URL }\fi
\providecommand{\bibinfo}[2]{#2}
\providecommand{\eprint}[2][]{\url{#2}}


\bibitem{DiValentino:2021izs}
E.~Di Valentino, O.~Mena, S.~Pan, L.~Visinelli, W.~Yang, A.~Melchiorri, D.~F.~Mota, A.~G.~Riess and J.~Silk,
Class. Quant. Grav. \textbf{38}, no.15, 153001 (2021)
doi:10.1088/1361-6382/ac086d
[arXiv:2103.01183 [astro-ph.CO]].

\bibitem{Kamionkowski:2022pkx}
M.~Kamionkowski and A.~G.~Riess,
Ann. Rev. Nucl. Part. Sci. \textbf{73}, 153-180 (2023)
doi:10.1146/annurev-nucl-111422-024107
[arXiv:2211.04492 [astro-ph.CO]].

\bibitem{Wang:2023bxf}
B.~Wang, M.~L{\'o}pez-Corredoira and J.~J.~Wei,
Mon. Not. Roy. Astron. Soc. \textbf{527}, no.3, 7692-7700 (2023)
doi:10.1093/mnras/stad3724
[arXiv:2311.18443 [astro-ph.CO]].

\bibitem{Verde:2023lmm}
L.~Verde, N.~Sch{\"o}neberg and H.~Gil-Mar{\'\i}n,
Ann. Rev. Astron. Astrophys. \textbf{62}, no.1, 287-331 (2024)
doi:10.1146/annurev-astro-052622-033813
[arXiv:2311.13305 [astro-ph.CO]].


\bibitem{Planck:2018vyg}
N.~Aghanim \textit{et al.} [Planck],
Astron. Astrophys. \textbf{641}, A6 (2020)
[erratum: Astron. Astrophys. \textbf{652}, C4 (2021)]
doi:10.1051/0004-6361/201833910
[arXiv:1807.06209 [astro-ph.CO]].

\bibitem{KiDS:2020suj}
M.~Asgari \textit{et al.} [KiDS],
Astron. Astrophys. \textbf{645}, A104 (2021)
doi:10.1051/0004-6361/202039070
[arXiv:2007.15633 [astro-ph.CO]].

\bibitem{DES:2021wwk}
T.~M.~C.~Abbott \textit{et al.} [DES],
Phys. Rev. D \textbf{105}, no.2, 023520 (2022)
doi:10.1103/PhysRevD.105.023520
[arXiv:2105.13549 [astro-ph.CO]].

\bibitem{Sugiyama:2023fzm}
S.~Sugiyama, H.~Miyatake, S.~More, X.~Li, M.~Shirasaki, M.~Takada, Y.~Kobayashi, R.~Takahashi, T.~Nishimichi and A.~J.~Nishizawa, \textit{et al.}
Phys. Rev. D \textbf{108}, no.12, 123521 (2023)
doi:10.1103/PhysRevD.108.123521
[arXiv:2304.00705 [astro-ph.CO]].


\bibitem{Flambaum:2019cih}
V.~V.~Flambaum and I.~B.~Samsonov,
Phys. Rev. D \textbf{100}, no.6, 063541 (2019)
doi:10.1103/PhysRevD.100.063541
[arXiv:1908.09432 [astro-ph.CO]].

\bibitem{Jedamzik:2020zmd}
K.~Jedamzik, L.~Pogosian and G.~B.~Zhao,
Commun. in Phys. \textbf{4}, 123 (2021)
doi:10.1038/s42005-021-00628-x
[arXiv:2010.04158 [astro-ph.CO]].

\bibitem{Gonzalez:2020fdy}
M.~Gonzalez, M.~P.~Hertzberg and F.~Rompineve,
JCAP \textbf{10}, 028 (2020)
doi:10.1088/1475-7516/2020/10/028
[arXiv:2006.13959 [astro-ph.CO]].

\bibitem{Ye:2021iwa}
G.~Ye, J.~Zhang and Y.~S.~Piao,
Phys. Lett. B \textbf{839}, 137770 (2023)
doi:10.1016/j.physletb.2023.137770
[arXiv:2107.13391 [astro-ph.CO]].

\bibitem{Agrawal:2019lmo}
P.~Agrawal, F.~Y.~Cyr-Racine, D.~Pinner and L.~Randall,
Phys. Dark Univ. \textbf{42}, 101347 (2023)
doi:10.1016/j.dark.2023.101347
[arXiv:1904.01016 [astro-ph.CO]].


\bibitem{Baldeschi:1983mq}
M.~R.~Baldeschi, R.~Ruffini and G.~B.~Gelmini,
Phys. Lett. B \textbf{122}, 221-224 (1983)
doi:10.1016/0370-2693(83)90688-3

\bibitem{Sin:1992bg}
S.~J.~Sin,
Phys. Rev. D \textbf{50}, 3650-3654 (1994)
doi:10.1103/PhysRevD.50.3650
[arXiv:hep-ph/9205208 [hep-ph]].

\bibitem{Lee:1995af}
J.~w.~Lee and I.~g.~Koh,
Phys. Rev. D \textbf{53}, 2236-2239 (1996)
doi:10.1103/PhysRevD.53.2236
[arXiv:hep-ph/9507385 [hep-ph]].

\bibitem{Matos:1998vk}
T.~Matos and F.~S.~Guzman,
Class. Quant. Grav. \textbf{17}, L9-L16 (2000)
doi:10.1088/0264-9381/17/1/102
[arXiv:gr-qc/9810028 [gr-qc]].


\bibitem{Hu:2000ke}
W.~Hu, R.~Barkana and A.~Gruzinov,
Phys. Rev. Lett. \textbf{85}, 1158-1161 (2000)
doi:10.1103/PhysRevLett.85.1158
[arXiv:astro-ph/0003365 [astro-ph]].

\bibitem{Boehmer:2007um}
C.~G.~Boehmer and T.~Harko,
JCAP \textbf{06}, 025 (2007)
doi:10.1088/1475-7516/2007/06/025
[arXiv:0705.4158 [astro-ph]].

\bibitem{Marsh:2010wq}
D.~J.~E.~Marsh and P.~G.~Ferreira,
Phys. Rev. D \textbf{82}, 103528 (2010)
doi:10.1103/PhysRevD.82.103528
[arXiv:1009.3501 [hep-ph]].

\bibitem{Hlozek:2014lca}
R.~Hlozek, D.~Grin, D.~J.~E.~Marsh and P.~G.~Ferreira,
Phys. Rev. D \textbf{91}, no.10, 103512 (2015)
doi:10.1103/PhysRevD.91.103512
[arXiv:1410.2896 [astro-ph.CO]].


\bibitem{Irsic:2017yje}
V.~Ir{\v{s}}i{\v{c}}, M.~Viel, M.~G.~Haehnelt, J.~S.~Bolton and G.~D.~Becker,
Phys. Rev. Lett. \textbf{119}, no.3, 031302 (2017)
doi:10.1103/PhysRevLett.119.031302
[arXiv:1703.04683 [astro-ph.CO]].

\bibitem{Armengaud:2017nkf}
E.~Armengaud, N.~Palanque-Delabrouille, C.~Y{\`e}che, D.~J.~E.~Marsh and J.~Baur,
Mon. Not. Roy. Astron. Soc. \textbf{471}, no.4, 4606-4614 (2017)
doi:10.1093/mnras/stx1870
[arXiv:1703.09126 [astro-ph.CO]].

\bibitem{Rogers:2020ltq}
K.~K.~Rogers and H.~V.~Peiris,
Phys. Rev. Lett. \textbf{126}, no.7, 071302 (2021)
doi:10.1103/PhysRevLett.126.071302
[arXiv:2007.12705 [astro-ph.CO]].


\bibitem{Park:2022lel}
S.~Park, D.~Bak, J.~W.~Lee and I.~Park,
JCAP \textbf{12}, 033 (2022)
doi:10.1088/1475-7516/2022/12/033
[arXiv:2207.07192 [astro-ph.CO]].

\bibitem{Koo:2023gfm}
H.~Koo, D.~Bak, I.~Park, S.~E.~Hong and J.~W.~Lee,
Phys. Lett. B \textbf{856}, 138908 (2024)
doi:10.1016/j.physletb.2024.138908
[arXiv:2311.03412 [astro-ph.GA]].

\bibitem{Bromley:2023yfi}
B.~C.~Bromley, P.~Sandick and B.~Shams Es Haghi,
Phys. Rev. D \textbf{110}, no.2, 023517 (2024)
doi:10.1103/PhysRevD.110.023517
[arXiv:2311.18013 [astro-ph.GA]].




\bibitem{Hui:2016ltb}
L.~Hui, J.~P.~Ostriker, S.~Tremaine and E.~Witten,
Phys. Rev. D \textbf{95}, no.4, 043541 (2017)
doi:10.1103/PhysRevD.95.043541
[arXiv:1610.08297 [astro-ph.CO]].

\bibitem{Ferreira:2020fam}
E.~G.~M.~Ferreira,
Astron. Astrophys. Rev. \textbf{29}, no.1, 7 (2021)
doi:10.1007/s00159-021-00135-6
[arXiv:2005.03254 [astro-ph.CO]].

\bibitem{Matos:2023usa}
T.~Matos, L.~A.~Ure{\~n}a-L{\'o}pez and J.~W.~Lee,
Front. Astron. Space Sci. \textbf{11}, 1347518 (2024)
doi:10.3389/fspas.2024.1347518
[arXiv:2312.00254 [astro-ph.CO]].


\bibitem{Lee:2025roi}
J.~W.~Lee,
[arXiv:2502.11568 [astro-ph.CO]].

\bibitem{Lopez-Sanchez:2025uwk}
J.~N.~L{\'o}pez-S{\'a}nchez,
[arXiv:2512.24309 [astro-ph.CO]].

\bibitem{Li:2013nal}
B.~Li, T.~Rindler-Daller and P.~R.~Shapiro,
Phys. Rev. D \textbf{89}, no.8, 083536 (2014)
doi:10.1103/PhysRevD.89.083536
[arXiv:1310.6061 [astro-ph.CO]].

\bibitem{Hartman:2021upg}
S.~T.~H.~Hartman, H.~A.~Winther and D.~F.~Mota,
JCAP \textbf{02}, no.02, 005 (2022)
doi:10.1088/1475-7516/2022/02/005
[arXiv:2108.07496 [astro-ph.CO]].


\bibitem{Urena-Lopez:2015gur}
L.~A.~Ure{\~n}a-L{\'o}pez and A.~X.~Gonzalez-Morales,
JCAP \textbf{07}, 048 (2016)
doi:10.1088/1475-7516/2016/07/048
[arXiv:1511.08195 [astro-ph.CO]].

\bibitem{Cookmeyer:2019rna}
T.~Cookmeyer, J.~Cookmeyer, D.~Grin and T.~L.~Smith,
Phys. Rev. D \textbf{101}, no.2, 023501 (2020)
doi:10.1103/PhysRevD.101.023501
[arXiv:1909.11094 [astro-ph.CO]].

\bibitem{Passaglia:2022bcr}
S.~Passaglia and W.~Hu,
Phys. Rev. D \textbf{105}, no.12, 123529 (2022)
doi:10.1103/PhysRevD.105.123529
[arXiv:2201.10238 [astro-ph.CO]].

\bibitem{Liu:2024yne}
R.~Liu, W.~Hu and D.~Grin,
Phys. Rev. D \textbf{112}, no.2, 023513 (2025)
doi:10.1103/1z4c-1w7f
[arXiv:2412.15192 [astro-ph.CO]].

\bibitem{Moss:2025ymr}
A.~Moss, L.~Gaughan and A.~M.~Green,
Phys. Rev. D \textbf{111}, no.12, 123530 (2025)
doi:10.1103/fy1y-rjst
[arXiv:2501.13662 [astro-ph.CO]].



\bibitem{Lesgourgues:2011re}
J.~Lesgourgues,
[arXiv:1104.2932 [astro-ph.IM]].

\bibitem{Blas:2011rf}
D.~Blas, J.~Lesgourgues and T.~Tram,
JCAP \textbf{07}, 034 (2011)
doi:10.1088/1475-7516/2011/07/034
[arXiv:1104.2933 [astro-ph.CO]].


\bibitem{ACT:2023kun}
M.~S.~Madhavacheril \textit{et al.} [ACT],
Astrophys. J. \textbf{962}, no.2, 113 (2024)
doi:10.3847/1538-4357/acff5f
[arXiv:2304.05203 [astro-ph.CO]].

\bibitem{Diez-Tejedor:2014naa}
A.~Diez-Tejedor, A.~X.~Gonzalez-Morales and S.~Profumo,
Phys. Rev. D \textbf{90}, no.4, 043517 (2014)
doi:10.1103/PhysRevD.90.043517
[arXiv:1404.1054 [astro-ph.GA]].

\bibitem{Boudon:2022dxi}
A.~Boudon, P.~Brax and P.~Valageas,
Phys. Rev. D \textbf{106}, no.4, 043507 (2022)
doi:10.1103/PhysRevD.106.043507
[arXiv:2204.09401 [astro-ph.CO]].

\bibitem{Cembranos:2018ulm}
J.~A.~R.~Cembranos, A.~L.~Maroto, S.~J.~N{\'u}{\~n}ez Jare{\~n}o and H.~Villarrubia-Rojo,
JHEP \textbf{08}, 073 (2018)
doi:10.1007/JHEP08(2018)073
[arXiv:1805.08112 [astro-ph.CO]].

\bibitem{Lee:2026udmclass}
S.~Lee,
\emph{UDMCLASS patch and reproducibility package},
version 1.0.0, Zenodo (2026),
\href{https://doi.org/10.5281/zenodo.22209422}{doi:10.5281/zenodo.22209422}.





\end{thebibliography}
\end{document}